\documentclass[pra,showpacs]{revtex4}
\usepackage{graphicx}
\usepackage{bm}

\begin{document}

\title{Interaction-free measurement with an imperfect absorber}

\author{Hiroo Azuma}
\altaffiliation{On leave from
Research Center for Quantum Information Science,
Tamagawa University Research Institute,
6-1-1 Tamagawa-Gakuen, Machida-shi, Tokyo 194-8610, Japan}
\email[Electronic address: ]{hiroo.azuma@m3.dion.ne.jp}
\affiliation{2-1-12 MinamiFukunishi-cho,
Oe, NishiKyo-ku, Kyoto-shi, Kyoto 610-1113, Japan}

\date{\today}

\begin{abstract}
In this paper, we consider interaction-free measurement (IFM) with imperfect interaction.
In the IFM proposed by Kwiat {\it et al}.,
we assume that interaction between an absorbing object and a probe photon is imperfect,
so that the photon is absorbed
with probability $1-\eta$ ($0\leq\eta\leq 1$) and it passes by the object without being absorbed
with probability $\eta$ when it approaches close to the object.
We derive the success probability $P$
that we can find the object without the photon absorbed
under the imperfect interaction as a power series in $1/N$,
and show the following result:
Even if the interaction between the object and the photon is imperfect,
we can let the success probability $P$ of the IFM get close to unity arbitrarily
by making the reflectivity of the beam splitter larger
and increasing the number of the beam splitters.
Moreover,
we obtain an approximating equation of $P$ for large $N$
from the derived power series in $1/N$.
\end{abstract}


\maketitle

\section{\label{section-introduction}Introduction}
In 1981, Dicke proposed a concept of interaction-free measurement (IFM)
\cite{Dicke}.
However, current discussion of IFM appears from the following problem
stated by Elitzur and Vaidman:
``Let us assume there is an object that absorbs a photon with strong interaction
if the photon approaches the object closely enough.
Can we examine whether or not the object exists without its absorption?''
\cite{Elitzur-Vaidman}.
The reason that we do not want to let the object absorb the photon is
that it might lead to an explosion, for example.
Elitzur and Vaidman themselves present a method of the IFM
that is inspired by the Mach-Zehnder interferometer.
Then a more refined one is proposed by Kwiat {\it et al}.
\cite{Kwiat-Weinfurter-1}.
An experiment of their IFM is reported in Ref.~\cite{Kwiat-White}.
The IFM finds wide application in quantum information processing
(the Bell-basis measurement, quantum computation, and so on)
\cite{Azuma-3,Azuma-4}.

According to the IFM proposed by Kwiat {\it et al}.,
the absorbing object is put in the interferometer that consists of $N$ beam splitters,
and we inject a photon into it to examine whether or not the object exists.
The probability that we can find the object in the interferometer arrives at unity
under the limit of $N\rightarrow\infty$ in the case where the interaction
between the object and the photon is strong enough and perfect.

In this paper, we consider the IFM of Kwiat {\it et al}. with imperfect interaction.
In ordinary IFM, the absorbing object is expected to absorb a photon
with probability unity when the photon approaches the object closely enough.
However, in this paper, we assume that the photon is absorbed with probability $1-\eta$ ($0\leq\eta\leq 1$)
and it passes by the object without being absorbed with probability $\eta$
when it approaches close to the object.
We estimate the success probability $P$ of the IFM,
namely the probability that we can find the object without the photon absorbed,
under this assumption.

This problem has been investigated in Ref.~\cite{Azuma-3} already.
In Ref.~\cite{Azuma-3}, although a correct approximating equation
of the success probability of the IFM
with the imperfect interaction is derived,
its derivation is wrong.
Hence, we give a right treatment of this problem in this paper.

\begin{figure}
\includegraphics[scale=0.65]{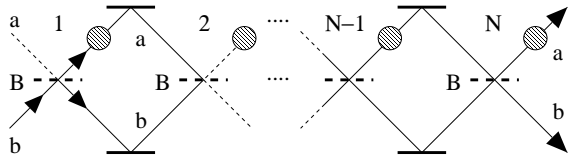}
\caption{\label{KWinterferometer}Interferometer of Kwiat {\it et al}.
for the IFM.}
\end{figure}

In the rest of this section, we give a short review of the IFM proposed by Kwiat {\it et al}.
They consider an interferometer that consists of $N$ beam splitters
as shown in Fig.~\ref{KWinterferometer}.
We describe the upper paths as $a$
and lower paths as $b$,
so that the beam splitters form the boundary line
between the paths $a$ and the paths $b$
in the interferometer.
We write a state with one photon on the paths $a$
as $|1\rangle_{a}$
and a state with no photon on the paths $a$ as $|0\rangle_{a}$.
This notation applies to the paths $b$ as well.
The beam splitter $B$ in Fig.~\ref{KWinterferometer}
works as follows:
\begin{equation}
B:
\left\{
\begin{array}{rrr}
|1\rangle_{a}|0\rangle_{b} & \rightarrow &
\cos\theta|1\rangle_{a}|0\rangle_{b}-\sin\theta|0\rangle_{a}|1\rangle_{b},\\
|0\rangle_{a}|1\rangle_{b} & \rightarrow &
\sin\theta|1\rangle_{a}|0\rangle_{b}+\cos\theta|0\rangle_{a}|1\rangle_{b}.
\end{array}
\right.
\label{definition-beam-splitter-B}
\end{equation}
[The transmissivity of $B$ is given by $T=\sin^{2}\theta$,
and the reflectivity of $B$ is given by $R=\cos^{2}\theta$
in Eq.~(\ref{definition-beam-splitter-B}).]

Let us throw a photon into the lower left port of $b$
in Fig.~\ref{KWinterferometer}.
If there is no object on the paths,
the wave function of the photon that comes from the $k$th
beam splitter is given by
\begin{eqnarray}
&&
\sin k\theta|1\rangle_{a}|0\rangle_{b}
+\cos k\theta|0\rangle_{a}|1\rangle_{b} \nonumber \\
&&\quad\quad
\mbox{for $k=0,1,...,N$.}
\end{eqnarray}
If we assume $\theta=\pi/2N$,
the photon that comes from the $N$th beam splitter goes to
the upper right port of $a$ with probability unity.

Next, we consider the case where there is an object that absorbs
the photon on the paths $a$.
We assume that the object is put on every path $a$ that comes
from each beam splitter,
and all of these $N$ objects are the same one.
The photon thrown into the lower left port of $b$
cannot go to the upper right port of $a$
because the object absorbs it.
If the incident photon goes to the lower right port
of $b$, it has not passed through paths $a$ in the interferometer.

Therefore, the probability that the photon goes
to the lower right port of $b$ is equal to the product
of the reflectivities of the beam splitters.
It is given by $P=\cos^{2N}\theta$.
In the limit of $N\rightarrow\infty$,
$P$ approaches $1$ as follows:
\begin{eqnarray}
\lim_{N\rightarrow\infty}P
&=&\lim_{N\rightarrow\infty}\cos^{2N}(\frac{\pi}{2N}) \nonumber \\
&=&\lim_{N\rightarrow\infty}
[1-\frac{\pi^{2}}{4N}+O(\frac{1}{N^{2}})] \nonumber \\
&=&1.
\end{eqnarray}

From the above discussion,
we can conclude that the interferometer of Kwiat et al.
directs an incident photon from the lower left port of $b$
with probability $P$ at least as follows:
(1) if there is no absorbing object in the interferometer,
the photon goes to the upper right port of $a$, and
(2) if there is the absorbing object in the interferometer,
the photon goes to the lower right port of $b$.
Furthermore, if we take large $N$, we can set $P$ arbitrarily
close to $1$.
Therefore, we can examine whether or not the object exists in the interferometer.

\section{\label{section-IFM-imperfect-interaction}
The IFM with imperfect interaction}
The IFM introduced in the former section is realized
by beam splitters and interaction between the absorbing object and the photon.
In this section,
we consider the case where the interaction is not perfect.
(We regard the beam splitters as accurate enough.)
We assume that the photon is absorbed
with probability $1-\eta$
and it passes by the object without being absorbed
with probability $\eta$
when it approaches close to the object.
We estimate the success probability $P$ of the IFM
under these assumptions.

We assume the following transformation
in Fig.~\ref{KWinterferometer}.
The photon that comes from each beam splitter to the upper path $a$ suffers
\begin{equation}
|\bar{0}\rangle
\rightarrow
\sqrt{\eta}|\bar{0}\rangle
+\sqrt{1-\eta}|\mbox{absorption}\rangle,
\label{imperfect-absorption-process}
\end{equation}
where $0\leq\eta\leq 1$ and $|\bar{0}\rangle=|1\rangle_{a}|0\rangle_{b}$.
$|\mbox{absorption}\rangle$ is the state where the object absorbs the photon.
We assume that it is normalized and orthogonal to
$\{|\bar{0}\rangle,|\bar{1}\rangle\}$,
where
$|\bar{0}\rangle=|1\rangle_{a}|0\rangle_{b}$
and
$|\bar{1}\rangle=|0\rangle_{a}|1\rangle_{b}$.

From now on, for simplicity,
we describe the transformations that are applied to the photon
as matrices in the basis $\{|\bar{0}\rangle,|\bar{1}\rangle\}$.
Writing
\begin{eqnarray}
|\bar{0}\rangle
&=&
|1\rangle_{a}|0\rangle_{b}
=
\left(
\begin{array}{c}
1 \\
0
\end{array}
\right), \nonumber \\
|\bar{1}\rangle
&=&
|0\rangle_{a}|1\rangle_{b}
=
\left(
\begin{array}{c}
0 \\
1
\end{array}
\right),
\label{definition-basis-vectors}
\end{eqnarray}
we can describe the beam splitter $B$ defined
in Eq.~(\ref{definition-beam-splitter-B}) as
\begin{equation}
B=
\left(
\begin{array}{cc}
\cos\theta & \sin\theta \\
-\sin\theta & \cos\theta
\end{array}
\right),
\label{definition-matrix-B}
\end{equation}
where $\theta=\pi/2N$
and the absorption process defined
in Eq.~(\ref{imperfect-absorption-process}) as
\begin{equation}
A=
\left(
\begin{array}{cc}
\sqrt{\eta} & 0           \\
0           & 1
\end{array}
\right),
\label{definition-matrix-A}
\end{equation}
where $0\leq\eta\leq 1$.
The matrix $A$ is not unitary because the process
defined by Eq.~(\ref{imperfect-absorption-process})
causes absorption of the photon (dissipation or decoherence).

\begin{figure}
\includegraphics[scale=0.8]{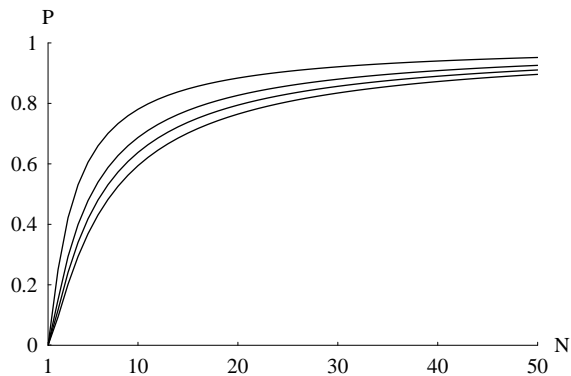}
\caption{\label{imperfectIFM-N-exactP}
Results of numerical calculation of
the success probability $P(N,\eta)$
from
Eqs.~(\ref{definition-basis-vectors}), (\ref{definition-matrix-B}),
(\ref{definition-matrix-A}),
and (\ref{definition-fidelity-IFMgate}).
With fixing $\eta$, we plot $P(N,\eta)$
as a function of $N$ and link them together by solid lines.
$\eta$ is the rate at which the object fails to absorb the photon.
$P$ and $\eta$ are dimensionless quantities.
$N$ is the number of the beam splitters.
Four cases of $\eta=0$, $0.05$, $0.1$, and $0.15$
are shown in order from top to bottom
as solid curves.}
\end{figure}

The probability that an incident photon from the lower left port of $b$
passes through the $N$ beam splitters and
is detected in the lower right port of $b$ in Fig.~\ref{KWinterferometer}
is given by
\begin{equation}
P(N,\eta)=|\langle\bar{1}|(BA)^{N-1}B|\bar{1}\rangle|^{2}.
\label{definition-fidelity-IFMgate}
\end{equation}
We plot results of numerical calculations of
the success probability $P(N,\eta)$
defined by Eqs.~(\ref{definition-basis-vectors}), (\ref{definition-matrix-B}),
(\ref{definition-matrix-A}),
and (\ref{definition-fidelity-IFMgate}) in Fig.~\ref{imperfectIFM-N-exactP}.
With fixing $\eta$, we plot $P(N,\eta)$ as a function of $N$
and link them together by solid lines.
In Fig.~\ref{imperfectIFM-N-exactP},
the four cases of $\eta=0$, $0.05$, $0.1$, and $0.15$
are shown in order from top to bottom.

Seeing Fig.~\ref{imperfectIFM-N-exactP},
we have the following question.
If $\eta\neq 0$, at what value does
$P(N,\eta)$ converge in the limit of $N\rightarrow\infty$?
At first glance,
$\lim_{N\rightarrow\infty}P(N,\eta)$ seems to depend on $\eta$.
However, Fig.~\ref{imperfectIFM-N-exactP} lets us expect
$\lim_{N\rightarrow\infty}P(N,\eta)=1$ $\forall\eta$.
In the next section, we show that this expectation is true.

\section{\label{section-Evaluation-success-probability}
Evaluation of the success probability}
In this section, we examine $P(N,\eta)$ defined in Eq.~(\ref{definition-fidelity-IFMgate})
for large $N$.
For this purpose, first we derive an exact formula of $P(N,\eta)$,
and second we expand $P(N,\eta)$ in powers of $1/N$ with fixing $\eta$.

First we derive the exact formula of $P(N,\eta)$.
We note the following.
Eigenvalues of the matrices $A$ and $B$ are given by $1$ and $\sqrt{\eta}$.
Thus, $\lim_{N\rightarrow\infty}P(N,\eta)$ never diverges to infinity.
We let $BA$ to be an upper triangular matrix $D$ by unitary transformation $U$ as follows:
\begin{equation}
D
=
\left(
\begin{array}{cc}
x & y \\
0 & z
\end{array}
\right)
=
U^{\dagger}BAU,
\end{equation}
where
\begin{eqnarray}
U
=
\left(
\begin{array}{cc}
U_{00} & U_{01} \\
U_{10} & U_{11}
\end{array}
\right),
\label{definition-matrix-U}
\end{eqnarray}
\begin{eqnarray}
U_{00}
&=&
\frac{(1-\sqrt{\eta})\cos\theta+r}{\sqrt{s}}, \nonumber\\
U_{01}
&=&
\frac{(1-\sqrt{\eta})\cos\theta-r}{\sqrt{t}}, \nonumber \\
U_{10}
&=&
\frac{2\sqrt{\eta}\sin\theta}{\sqrt{s}}, \nonumber \\
U_{11}
&=&
-\frac{2\sin\theta}{\sqrt{t}},
\end{eqnarray}
\begin{eqnarray}
r
&=&
\sqrt{(1-\sqrt{\eta})^{2}\cos^{2}\theta-4\sqrt{\eta}\sin^{2}\theta}, \nonumber \\
s
&=&
4\eta\sin^{2}\theta+[(1-\sqrt{\eta})\cos\theta+r]^{2}, \nonumber \\
t
&=&
4\sin^{2}\theta+[(1-\sqrt{\eta})\cos\theta-r]^{2},
\end{eqnarray}
and
\begin{eqnarray}
x
&=&
\frac{1}{2}[(1+\sqrt{\eta})\cos\theta-r], \nonumber \\
y
&=&
-\frac{1}{\sqrt{st}}(1-\sqrt{\eta})\sin^{2}\theta
[2(1+\sqrt{\eta})\cos\theta \nonumber \\
&&\quad
+\sqrt{2}
\sqrt{1-6\sqrt{\eta}+\eta+(1+\sqrt{\eta})^{2}\cos 2\theta}], \nonumber \\
z
&=&
\frac{1}{2}[(1+\sqrt{\eta})\cos\theta+r].
\end{eqnarray}
We obtain $D^{N-1}$ by induction as follows:
\begin{equation}
D^{N-1}=
\left(
\begin{array}{cc}
X & Y \\
0 & Z
\end{array}
\right),
\label{definition-matrix-D-N-1}
\end{equation}
where
\begin{eqnarray}
X
&=&
x^{N-1}, \nonumber \\
Y
&=&
y(x^{N-2}+z^{N-2})+xyz\frac{x^{N-3}-z^{N-3}}{x-z}, \nonumber \\
Z
&=&
z^{N-1}.
\label{exact-components-D-N-1}
\end{eqnarray}
From the above calculations, we obtain the exact formula of $P(N,\eta)$ as
\begin{equation}
P(N,\eta)=|\langle\bar{1}|UD^{N-1}U^{\dagger}B|\bar{1}\rangle|^{2}.
\label{exact-formula-fidelity-IFMgate}
\end{equation}

Second we expand $P(N,\eta)$ in powers of $1/N$ with fixing $\eta$.
(We note $\theta=\pi/2N$.)
We can expand components of the matrices $B$, $U$, and $D$  in powers of $1/N$
as follows:
\begin{eqnarray}
\cos\theta
&=&
1-\frac{\pi^{2}}{8N^{2}}+O(\frac{1}{N^{4}}), \nonumber \\
\sin\theta
&=&
\frac{\pi}{2N}(1-\frac{\pi^{2}}{24N^{2}})+O(\frac{1}{N^{5}}),
\label{components-B-expansion}
\end{eqnarray}
\begin{eqnarray}
U_{00}
&=&
1-\frac{\pi^{2}}{8N^{2}}\frac{\eta}{(1-\sqrt{\eta})^{2}}+O(\frac{1}{N^{4}}), \nonumber \\
U_{01}
&=&
\frac{\pi}{2N}\frac{\sqrt{\eta}}{1-\sqrt{\eta}}
[1+\frac{\pi^{2}}{24N^{2}}\frac{2+2\sqrt{\eta}-\eta}{(1-\sqrt{\eta})^{2}}] \nonumber \\
&&\quad\quad
+O(\frac{1}{N^{5}}),\nonumber \\
U_{10}
&=&
\frac{\pi}{2N}\frac{\sqrt{\eta}}{1-\sqrt{\eta}}
[1+\frac{\pi^{2}}{24N^{2}}\frac{2-3\eta+\eta^{3/2}}{(1-\sqrt{\eta})^{3}}] \nonumber \\
&&\quad\quad
+O(\frac{1}{N^{5}}),\nonumber \\
U_{11}
&=&
-1+\frac{\pi^{2}}{8N^{2}}\frac{\eta}{(1-\sqrt{\eta})^{2}}+O(\frac{1}{N^{4}}),
\label{components-U-expansion}
\end{eqnarray}
and
\begin{eqnarray}
x
&=&
\sqrt{\eta}
[1+\frac{\pi^{2}}{8N^{2}}\frac{1-\eta}{(1-\sqrt{\eta})^{2}}]+O(\frac{1}{N^{4}}), \nonumber \\
y
&=&
-\frac{\pi}{2N}
[1-\frac{\pi^{2}}{24N^{2}}]+O(\frac{1}{N^{5}}), \nonumber \\
z
&=&
1-\frac{\pi^{2}}{8N^{2}}\frac{1-\eta}{(1-\sqrt{\eta})^{2}}+O(\frac{1}{N^{4}}).
\label{expansion-components-D}
\end{eqnarray}

Next, from Eqs.~(\ref{exact-components-D-N-1}) and (\ref{expansion-components-D}),
we expand components of $D^{N-1}$ in powers of $1/N$ with fixing $\eta$.
$X$ and $Z$ can be written as follows:
\begin{eqnarray}
X
&=&
\sqrt{\eta}^{N-1}
[1+\frac{\pi^{2}}{8N}\frac{1-\eta}{(1-\sqrt{\eta})^{2}}+O(\frac{1}{N^{2}})],
\label{expansion-component-X} \\
Z
&=&
1-\frac{\pi^{2}}{8N}\frac{1-\eta}{(1-\sqrt{\eta})^{2}}
+O(\frac{1}{N^{2}}).
\label{expansion-component-Z}
\end{eqnarray}
In Eq.~(\ref{expansion-component-X}), a factor $\sqrt{\eta}^{N}$ appears.
We never expand $\sqrt{\eta}^{N}$ in powers of $1/N$
and regard it as a constant.
Because $N$ is a large number, we can assume $\sqrt{\eta}^{N}\ll 1/N$.
Thus, we can write $Y$ in the form,
\begin{equation}
Y=
-\frac{\pi}{2N}\frac{1-\sqrt{\eta}^{N-1}}{1-\sqrt{\eta}}
[1-\frac{\pi^{2}}{8N}\frac{1+\sqrt{\eta}}{1-\sqrt{\eta}}]
+O(\frac{1}{N^{3}}).
\label{expansion-component-Y}
\end{equation}

Substituting Eqs.~(\ref{definition-matrix-B}),
(\ref{definition-matrix-U}),
(\ref{definition-matrix-D-N-1}),
(\ref{components-B-expansion}),
(\ref{components-U-expansion}),
(\ref{expansion-component-X}),
(\ref{expansion-component-Z}),
and
(\ref{expansion-component-Y})
into Eq.~(\ref{exact-formula-fidelity-IFMgate}),
we obtain an expansion of the success probability $P$ in powers of $1/N$
as follows:
\begin{equation}
P(N,\eta)=1-\frac{\pi^{2}}{4}\frac{1+\sqrt{\eta}}{1-\sqrt{\eta}}\frac{1}{N}
+O(\frac{1}{N^{2}}).
\label{expansion-success-probability}
\end{equation}
Hence, we can let $P$ get close to unity arbitrarily by increasing $N$.
[If we make $N$ larger, the reflectivity of the beam splitter
$R=\cos^{2}(\pi/2N)$ becomes larger and gets close to unity.]

\begin{figure}
\includegraphics[scale=0.8]{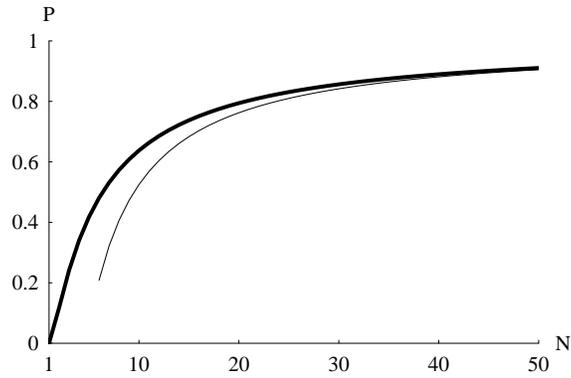}
\caption{\label{imperfectIFM-N-approxP}
Results of numerical calculation of the success probability $P$
as a function of $N$ with $\eta=0.1$.
$\eta$ is the rate at which the object fails to absorb the photon.
$P$ and $\eta$ are dimensionless quantities.
$N$ is the number of the beam splitters.
A thick solid curve represents an exact result from
Eqs.~(\ref{definition-basis-vectors}), (\ref{definition-matrix-B}),
(\ref{definition-matrix-A}),
and (\ref{definition-fidelity-IFMgate}).
A thin solid curve represents an approximate result
from Eq.~(\ref{approximating-eq-P}).}
\end{figure}

From Eq.~(\ref{expansion-success-probability}), we obtain the following
approximating equation of $P$ for large $N$,
\begin{equation}
P
\simeq
1-\frac{\pi^{2}}{4}\frac{1+\sqrt{\eta}}{1-\sqrt{\eta}}\frac{1}{N}.
\label{approximating-eq-P}
\end{equation}
In Fig.~\ref{imperfectIFM-N-approxP},
we plot results of numerical calculation of the success probability $P$
as a function of $N$ with $\eta=0.1$.
A thick solid curve represents an exact result from
Eqs.~(\ref{definition-basis-vectors}), (\ref{definition-matrix-B}),
(\ref{definition-matrix-A}),
and (\ref{definition-fidelity-IFMgate}).
A thin solid curve represents an approximate result
from Eq.~(\ref{approximating-eq-P}).
Seeing Fig.~\ref{imperfectIFM-N-approxP},
we find that Eq.~(\ref{approximating-eq-P}) is a good approximation to $P$
for large $N$.

\section{\label{section-conclusion}Conclusion}
We show that even if the interaction between the object and the photon is imperfect,
we can let the success probability $P$ of the IFM get close to unity arbitrarily
by making the reflectivity of the beam splitter larger
and increasing the number of the beam splitters.
We obtain an approximating equation of $P$ with the imperfect interaction.
To overcome the imperfection of the interaction,
we need to prepare a large number of beam splitters
and let their transmission rate get smaller.

\end{document}